\documentclass[12pt,preprint]{aastex}

\def\ltsima{$\; \buildrel < \over \sim \;$}
\def\gtsima{$\; \buildrel > \over \sim \;$}
\def\lsim{\lower.5ex\hbox{\ltsima}}
\def\gsim{\lower.5ex\hbox{\gtsima}}
\def\lapp{\ifmmode\stackrel{<}{_{\sim}}\else$\stackrel{<}{_{\sim}}$\fi}
\def\gapp{\ifmmode\stackrel{>}{_{\sim}}\else$\stackrel{<}{_{\sim}}$\fi}

\setbox0=\hbox{\rm0}
 
\shorttitle{Variable stars in Tezan~5}
\shortauthors{Origlia et al.}
 
\begin{document} 

\title{Variable stars in Terzan~5: additional evidence of multi-age and multi-iron stellar populations
\footnote{Based on observations collected at the 
Very Large Telescope of the European Southern Observatory under program 
097.D-0337. Also based on observations (GO 10845) with the NASA/ESA Hubble Space Telescope, 
obtained at the Space Telescope Science Institute, which is operated by AURA, Inc., under NASA contract NAS 5-26555.}}

\author{
L. Origlia\altaffilmark{2},
A. Mucciarelli\altaffilmark{3,2},
G. Fiorentino\altaffilmark{2}
F. R. Ferraro\altaffilmark{3,2},
E. Dalessandro\altaffilmark{2},
B. Lanzoni\altaffilmark{3,2},
R. M. Rich\altaffilmark{4},
D. Massari\altaffilmark{5},
R. R.  Contreras\altaffilmark{6,7},
N. Matsunaga\altaffilmark{8}
}
\affil{\altaffilmark{2} INAF-Osservatorio di Astrofisica e Scienza dello Spazio,Via Gobetti 93/3, I-40129 Bologna, Italy, livia.origlia@inaf.it}
\affil{\altaffilmark{3} Dipartimento di Fisica e Astronomia, Universit\`a degli Studi di Bologna, Via Gobetti 93/2, 40129 Bologna, Italy}
\affil{\altaffilmark{4} Department of Physics and Astronomy, University of California at Los Angeles, 430 Portola Plaza Box 951547, Los Angeles, CA 90095-1547, USA}
\affil{\altaffilmark{5} Kapteyn Astronomical Institute, University of Groningen, Groningen, The Netherlands}
\affil{\altaffilmark{6} Millennium Institute of Astrophysics, Santiago, Chile}
\affil{\altaffilmark{7} Instituto de Astrof\'{i}sica, Pontificia Universidad Cat\'olica de Chile, Av. Vicu\~na Mackenna 4860, 782-0436 Macul, Santiago, Chile}
\affil{\altaffilmark{8} Department of Astronomy, The University of Tokyo, 7-3-1 Hongo, Bunkyo-ku, Tokyo 113-0033, Japan}
\date{11 December 2018}

\begin{abstract}
Terzan~5 is a complex stellar system in the Galactic bulge, harboring
stellar populations with very different iron content ($\rm
\Delta[Fe/H]\sim1$ dex) and with ages differing by several Gyrs.  Here
we present an investigation of its variable stars. We report on the
discovery and characterization of three RR Lyrae stars.  For these
newly discovered RR Lyrae and for six Miras of known periods we
provide radial velocity and chemical abundances from spectra acquired
with X-SHOOTER at the VLT.  We find that the three RR Lyrae and the
three short period Miras (P$<$300 d) have radial velocity consistent
with being Terzan~5 members.  They have sub-solar iron abundances and
enhanced [$\alpha$/Fe], well matching the age and abundance patterns
of the 12 Gyr metal-poor stellar populations of Terzan~5.  Only one,
out of the three long period (P$>$300 d) Miras analyzed in this study,
has a radial velocity consistent with being Terzan~5 member. Its
super-solar iron abundance and solar-scaled [$\alpha$/Fe] nicely match
the chemical properties of the metal rich stellar population of
Terzan~5 and its derived mass nicely agrees with being several Gyrs
younger than the short period Miras. This young variable is an
additional proof of the surprising young sub-population discovered in
Terzan~5.
\end{abstract}


\section{Introduction}
\label{intro}
Terzan~5 (hereafter Ter5) is a stellar system commonly catalogued as a
globular cluster located in the bulge of the Milky Way (MW).  It is
affected by large \citep{ort96,bar98,val07} and differential
\citep{mas12} reddening, with an average color excess E(B-V)=2.38.
This stellar system also harbors an exceptionally large population of
milli-second pulsars (MSPs; \citealp{ransom05,cad18}. See also the
updated list at {\tt http://www.naic.edu/$\sim$pfreire/GCpsr.html})
and a proto-MSP \citep{fer15}.

Adaptive optics imaging with the VLT and near IR spectroscopy with
Keck revealed the presence of two distinct red clumps in the
color-magnitude diagram, that cannot be explained by differential
reddening or distance effects, while they show very different iron
abundances ([Fe/H]=$-$0.2 and +0.3 dex, respectively,
\citealp{fer09}).  Subsequent spectroscopic studies
\citep{ori11,ori13,mas14a,mas14b} fully confirmed this finding and
revealed an additional, minor (a few percent) stellar population (SP)
of metal poor stars at [Fe/H]$\sim-0.8$ dex, bringing the overall
metallicity range covered by the SPs of Ter5 to $\sim$1 dex. Note that
such a large iron spread has never been observed in any Galactic
globular cluster, with the only exception of $\omega$ Centauri in the
halo, which is now believed to be the remnant of a dwarf galaxy
accreted by the MW \citep[see e.g.][]{bekki03,bekki06}.  The sub-solar
SPs of Ter5, with peaks at [Fe/H]$\sim-0.2$ and $-$0.8 dex, are
$\alpha$-enhanced and they likely formed early and quickly from a gas
mainly polluted by type II supernovae (SNe).  The super-solar
component at [Fe/H]$\sim$+0.3 dex is more centrally concentrated than
the others \citep{fer09,lan10}, and it has approximately solar
[$\alpha$/Fe] ratio, requiring a progenitor gas polluted by both SNe
II and SNe Ia on a longer timescale.  Recently, by means of HST and
ground-based adaptive optics deep imaging, we detected two distinct
main-sequence turnoff points in Ter5, providing the age of the two
main stellar populations \citep{fer16}: 12 Gyr for the (dominant)
sub-solar component and 4.5 Gyr for the one at super-solar
metallicity.

An intriguing scenario is emerging from these observational facts.
({\it i}) Ter5 is not a genuine globular cluster, nor the result
of the merging of two globulars;
({\it ii}) it has experienced a complex star formation (SF) and chemical
enrichment history, possibly
characterized by
short SF episodes (thus accounting for
the small metallicity spread measured within each sub-population); 
({\it iii}) Ter5 was 
originally much more massive ($>$10$^7$ M$_{\odot}$) than today (2$\times $10$^6$ M$_{\odot}$, \citealp{lan10}), 
thus able to retain the SN ejecta and also to explain its huge population of MSPs;
({\it iv}) Ter5 seems to have formed and evolved in deep connection with the bulge \citep{mas15}.

Indeed, there is a striking chemical
similarity between the Ter5 and the bulge SPs, which
show a metallicity distribution with two major peaks at sub-solar and
super-solar [Fe/H] and a tail/minor peak towards lower metallicities
(see e.g. \citealp{zoc08,hil11,joh11,ric12,ness13a,ness13b,ben13}, 
and also \citealp{joh14,roj14,gon15,ryd16,jon17,sch17}). 
These bulge SPs also show [$\alpha$/Fe] enhancement up to
about solar metallicity, and then a progressive decline towards solar
[$\alpha$/Fe] at super-solar [Fe/H].

Among the mechanisms that could have contributed to form the Galactic bulge, early (gas \& stars) merging, 
friction of massive clumps, proto-disk evaporation etc. 
have been proposed since several years \citep[see e.g.][]{imm04,car07,elm08}.
More recently, it has been suggested that the giant clumps
observed in high redshift galaxies                  
\citep[see e.g.][]{gen11,tac15} could have originated by the clustering of smaller, seed clumps with
typical masses of $10^7$-10$^8$ M$_{\odot}$
\citep[see e.g.][]{beh16}, in a bottom-up scenario.
In this framework (see e.g. \citealp{fer16}), the proto-Ter5 could have been one of those seed clumps 
that did not grow and merge into the Galactic bulge, but 
for some unknown reasons evolved in isolation and self-enriched.
Very recently, \citet{mck18} suggested that the super-solar component of Ter5 could originated from the gas  
of a giant molecular cloud colliding with the proto-Ter5 some Gyrs ago.

An important and still unexplored tile of the Ter5 puzzle is its population of variable stars,
which are independent and powerful tracers of SP properties. Indeed, as well known, while RR Lyrae trace old ($>$10 Gyr) SPs,  
Mira stars with different pulsation periods and metallicity can  trace SPs of different ages. Short period Miras (P$<$300d) 
usually trace  old SPs, while long period  (P$>$300d) ones are normally younger.

\section{Searching for variables stars in Ter5}

\begin{deluxetable}{llllllllll}
\rotate
\tabletypesize{\footnotesize} \tablecaption{
Coordinates, stellar parameters and chemical abundances for the observed variable stars in Ter5.}
\tablewidth{0pt} \tablehead{ 
\colhead{header}& 
\colhead{RR1}&
\colhead{RR2}&
\colhead{RR3}&
\colhead{M1}&
\colhead{M2}&
\colhead{M3}&
\colhead{M4}&
\colhead{M5}&
\colhead{M6}}
\startdata
RA (h~m~s)    & ~17~48~02.8& ~17~48~08.2& ~17~48~04.3& ~17~47~59.5& ~17~48~09.3& ~17~48~07.2& ~17~48~03.4& ~17~47~54.3& ~17~47~53.2 \\
Dec ($^o$ ' ")& $-$24~47~47.5& $-$24~45~42.1& $-$24~47~37.7& $-$24~47~17.6& $-$24~47~06.3& $-$24~46~26.6& $-$24~46~42.0& $-$24~49~54.6& $-$24~44~34.0 \\
P (days)      & ~0.72& ~0.64& ~0.89& ~217& ~269& ~261& ~464& ~377& ~455 \\
Phase$^a$ & ~0.70 & ~0.49 & ~0.63 & ~0.61 & ~0.38& ~0.51 & ~0.73 & ~0.42 & ~0.35 \\
$\rm T_{eff}$ & ~6250& ~6000& ~6000& ~3100& ~3400& ~3000& ~3000& ~3100& ~3200 \\
RV$^b$        & $-$77& $-$98& $-$92& $-$89 &$-$95& $-$75& $-$119& +162& +48 \\
$\rm [Fe/H]$  & $-$0.72$\pm$0.03& $-$0.71$\pm$0.03& $-$0.67$\pm$0.01& $-$0.27$\pm$0.01& $-$0.33$\pm$0.05& $-$0.26$\pm$0.07& +0.32$\pm$0.01& +0.27$\pm$0.01& +0.31$\pm$0.05 \\
$\rm [Ca/Fe]$ & +0.27$\pm$0.15& +0.37$\pm$0.04& +0.35$\pm$0.03& +0.31$\pm$0.15& +0.30$\pm$0.16& +0.32$\pm$0.17& +0.06$\pm$0.15& $-$0.04$\pm$0.06& $-$0.02$\pm$0.16 \\
$\rm [Si/Fe]$ & +0.39$\pm$0.16& +0.36$\pm$0.15& +0.37$\pm$0.03& +0.38$\pm$0.15& +0.40$\pm$0.16& +0.30$\pm$0.17& +0.02$\pm$0.03& $-$0.03$\pm$0.15& $-$0.05$\pm$0.06 \\
$\rm [Mg/Fe]$ & +0.35$\pm$0.15& +0.31$\pm$0.15& +0.36$\pm$0.03& +0.45$\pm$0.15& +0.36$\pm$0.09& +0.33$\pm$0.10& -0.00$\pm$0.15& $-$0.01$\pm$0.15& $-$0.01$\pm$0.16 \\
$\rm [Ti/Fe]$ & +0.22$\pm$0.15& +0.31$\pm$0.15& +0.29$\pm$0.15& +0.35$\pm$0.02& +0.25$\pm$0.05& +0.33$\pm$0.09& -0.06$\pm$0.03& $-$0.06$\pm$0.07& $-$0.02$\pm$0.16 \\
$\rm [Al/Fe]$ & +0.25$\pm$0.15& +0.28$\pm$0.15& +0.36$\pm$0.03& +0.36$\pm$0.03& +0.41$\pm$0.06& +0.41$\pm$0.07& -0.07$\pm$0.09& +0.13$\pm$0.02& $-$0.05$\pm$0.07 \\
$\rm [Na/Fe]$ & -0.07$\pm$0.15& +0.04$\pm$0.15& -0.07$\pm$0.15& -0.01$\pm$0.15& +0.01$\pm$0.16& +0.08$\pm$0.17& +0.00$\pm$0.03& +0.07$\pm$0.15& -0.01$\pm$0.16 \\
$\rm [Mn/Fe]$ & --            & --            & --            & +0.04$\pm$0.15& +0.05$\pm$0.16& -0.03$\pm$0.17& $-$0.07$\pm$0.15& $-$0.01$\pm$0.15& $-$0.02$\pm$0.16 \\
$\rm [K/Fe]$  & --            & --            & --            & $-$0.06$\pm$0.15& +0.02$\pm$0.19& -0.06$\pm$0.17& $-$0.02$\pm$0.15& $-$0.03$\pm$0.15& $-$0.07$\pm$0.16 \\
$\rm [O/Fe]$  & --            & --            & --            & +0.32$\pm$0.04& +0.41$\pm$0.12& +0.44$\pm$0.18& $-$0.08$\pm$0.12& $-$0.01$\pm$0.02& +0.03$\pm$0.10 \\
\enddata
\tablenotetext{a}{Approximative phase at the epoch of the spectroscopic observation with XSHOOTER.}
\tablenotetext{b}{Heliocentric radial velocity in $\rm km~s^{-1}$, with typical uncertainties of $\pm$2 km/s.}
\label{var}
\end{deluxetable}

\begin{figure*}
\begin{center}
\includegraphics[scale=0.15]{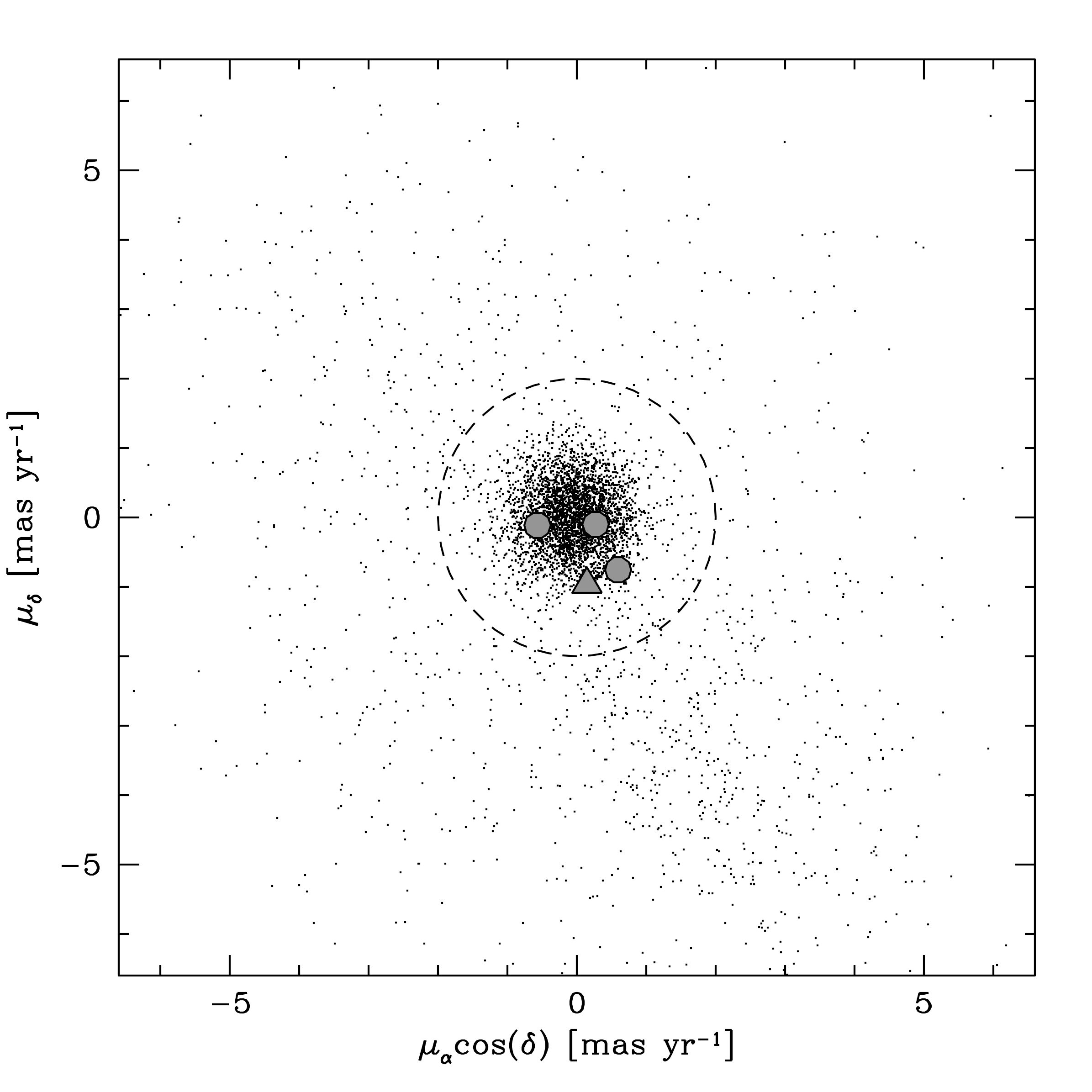}
\caption{Vector point diagram (VPD) of the HST proper motions measured
  by \citet{mas15} for stars with V$<$24 towards Ter5.  The three grey
  circles are the newly discovered RR Lyrae presented in this work
  (RR1, RR2 and RR3), while the grey triangle is RR4 \citep{edm01}.
  The dashed circle delimits the area in the VPD where likely Ter5
  member stars are distributed.  Clearly, all RR Lyrae have proper
  motions fully consistent with the mean motion of Ter5.}
\label{PM}
\end{center}
\end{figure*}  

\begin{figure*}
\begin{center}
\includegraphics[scale=0.15]{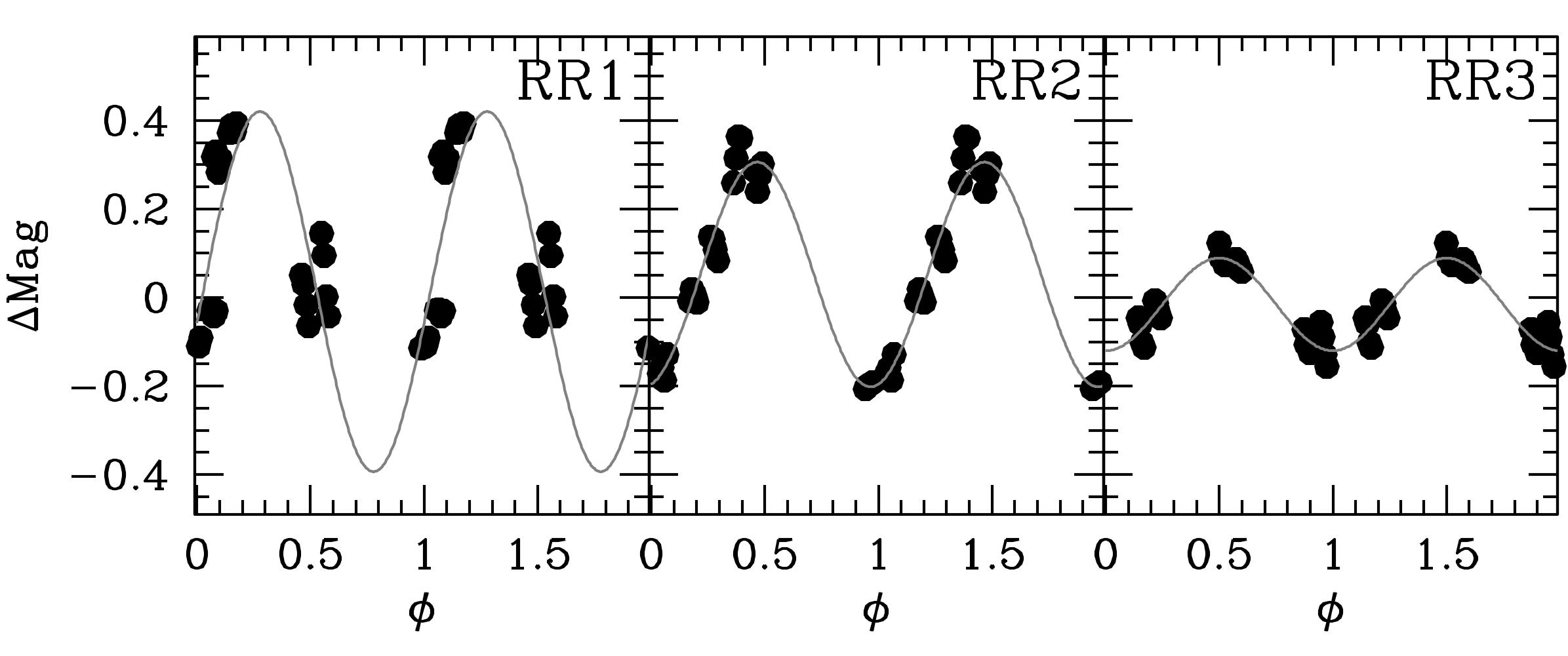}
\caption{Light curves from HST photometry of the three, newly
  discovered RR Lyrae in Ter5.}
\label{light_curves}
\end{center}
\end{figure*}  

A systematic search for RR Lyrae in Ter5 is still
missing. \citet{edm01} discovered one such candidate star by using
HST-NICMOS.  Recently, we used a sequence of 24 images acquired with
the HST Wide Field Planetary Camera 2 (WFPC2), obtained through
proposal GO-10845 (PI: Ferraro), to identify other candidates.  Each
image has an exposure time $t_{exp}=500$sec, 12 images have been
obtained in the F606W and 12 in the F814W passbands.

In this data-set the planetary camera (with the highest spatial
resolution of $\sim0.046\arcsec$ px$^{-1}$) is roughly centered on the
cluster core.  The photometric reduction was performed independently
for each image and chip by using {\it DAOPHOT} \citep{stetson87} and
following the approach adopted to study other dense stellar fields
\citep[e.g.][]{dal18}.  Briefly, dozens of bright and isolated stars
have been selected in each image to model the point-spread function.
A first star list was obtained for each image by independently fitting
all the star-like sources at $3 \sigma$ from the local background.  A
master-list including all stars detected in at least 13 (i.e. n/2+1)
images was created and a fit was then forced at the corresponding
positions in each frame by using {\it ALLFRAME} \citep{stetson94}.
For each image, instrumental magnitudes were corrected for charge
transfer efficiency by using the prescriptions described in
\citet{dolphin00}.  Different magnitude estimates were homogeneized by
using {\it DAOMATCH} and {\it DAOMASTER} and were then reported to the
VEGAMAG photometric system by following \citet{holtzman95} and related
zeropoints.  Instrumental coordinates were roto-translated to the
absolute coordinate system by using the stars in common with the HST
catalog used in \citet{fer09} as secondary astrometric standard and
the cross-correlation software {\textrm CataXcorr}.

By taking advantage of the relatively large number of images in this
data-set and the photometric quality of HST, we performed a detailed
variability analysis of stars with magnitudes in the range
$16.5<m_{\rm F814W}<18.5$ (corresponding to the red clump magnitude
level) looking for candidate RR Lyrae stars.  The analysis of variable
stars was carried out in the F606W and F814W bands, separately.  As a
first diagnostic to identify variables, we used the {\it variability
  indicator} provided by {\it DAOPHOT}.  We selected only stars
showing variability values significantly larger than those of the bulk
of stars with similar magnitudes in both bands.  Then, we checked
visually their preliminary light-curves and considered only stars
showing coherent evidence of variability in F606W and F814W.  We
identified in this way four candidate RR Lyrae.  It turns out that one
of them (RR4) corresponds to the candidate RR Lyrae star V1 found by
\citet{edm01} by using near IR HST imagery.  The other three are newly
discovered candidate variables.

All the four variables are located in the central $80\arcsec$ (where
the contamination by bulge field giants is negligible, of the order of
2\%; \citealp{mas14a,mas14b}). For these stars there are HST proper
motion estimates by \citet{mas15}, and Fig.~\ref{PM} shows the vector
point diagram with the location of Ter5 stars and the four known RR
Lyrae.  The variables are well clumped within the bulk of the Ter5
distribution, thus providing a robust evidence for their membership.
Because of the reddening, these variables are too faint to be measured
by Gaia.

We analysed the light curves of the three new candidate RR Lyrae by
using the Graphical Analyzer of Time Series ({\it
GRATIS2})\footnote{{\it GRATIS2} is a private software developed at
the Bologna Observatory by P. Montegriffo.}. It uses both the Lomb
periodogram \citep{lomb76} and the best fit of the data with a
truncated Fourier series \citep{barning63}.  The final periods adopted
to fold the light curves are those that minimize the rms scatter of
the truncated Fourier series that best fit the data.

To use the largest possible sample of data-points, for each variable
we have scaled the F814W magnitudes to the F606W ones, by using the
amplitude ratio A$_{\rm F606W}$/A$_{\rm F814W}$=1.49 derived by
\citet{fio12}. The Fourier analysis was then applied to the combined
F606W and F814W light-curves, shown in Fig.~\ref{light_curves}.  
For all the candidates RR Lyrae we
obtained periods in the 0.6-0.9 d range (see Table~\ref{var}), typical
of fundamental pulsators of AB type.  The derived light-curves are

However, we note that, 
given the incomplete sampling of the currently available light curves,  
particularly in the case of RR1,
the inferred pulsation periods and especially amplitudes 
should be considered as indicative.
More precise estimates will follow when better sampled light-curves 
will become available.

An extensive search for Miras in Ter5 was done by using the SIRIUS
near IR camera attached to the 1.4-m Infrared Survey Facility (IRSF)
telescope over the period from 2002 to 2005 (Matsunaga, 2007, PhD
thesis).  Among the detected Mira candidates, six are possibly Ter5
members according to their K-band magnitudes and period--luminosity
(P-L) relation (see Sloan et al. 2010 for a discussion of five such
candidates).  We note that these six Miras (M1 to M6) are known as V2,
V6, V8, V5, V7, and V12, respectively in the lists of variable stars
in Ter5 by Clement et al. (2001), Matsunaga (2007), and Sloan et
al. (2010).  Three Miras have periods P$<$300 d, while the other three
stars have P$>$300 d (see Table~\ref{var}).  Moreover, we note that
four (M1 to M4) out of these six Miras are located in the inner
$100\arcsec$, where field contamination by giant stars is negligible,
and the other two targets are at distances of about $200\arcsec$ from
the Ter5 center, where contamination is still reasonably low (about
30\%, \citealp{mas14a,mas14b}). 

These variables lack proper motions from \citet{mas15}, since they
were saturated in that survey, while they are sufficiently bright for
Gaia. First proper motion estimates from DR2 \citep{gaia18} indicate
membership for M1, M2 and M3 and non-membership for M5 and M6, while
no measurements are available for M4.

By combining observed and theoretical pulsation properties of Miras
with evolutionary models, \citet{fea96} derived the following relation
(his equation 15; see also \citealp{woo83}): log M/M$_{\odot}$ = 0.470
log P +0.356 [Fe/H] $-1.340$, that, for a given metallicity, predicts
larger masses (i.e., younger ages) for longer pulsation periods.
Interestingly, the mass difference between Miras with different
pulsation periods becomes even larger if the longer period variables
are also more metal-rich.  Thus, accurate determination of the metal
content of the Mira variables is key to constrain their mass (hence
their age) and the properties of their parent stellar population
\citep[e.g.][]{cat16}.

\begin{figure*}
\begin{center}
\includegraphics[scale=0.15]{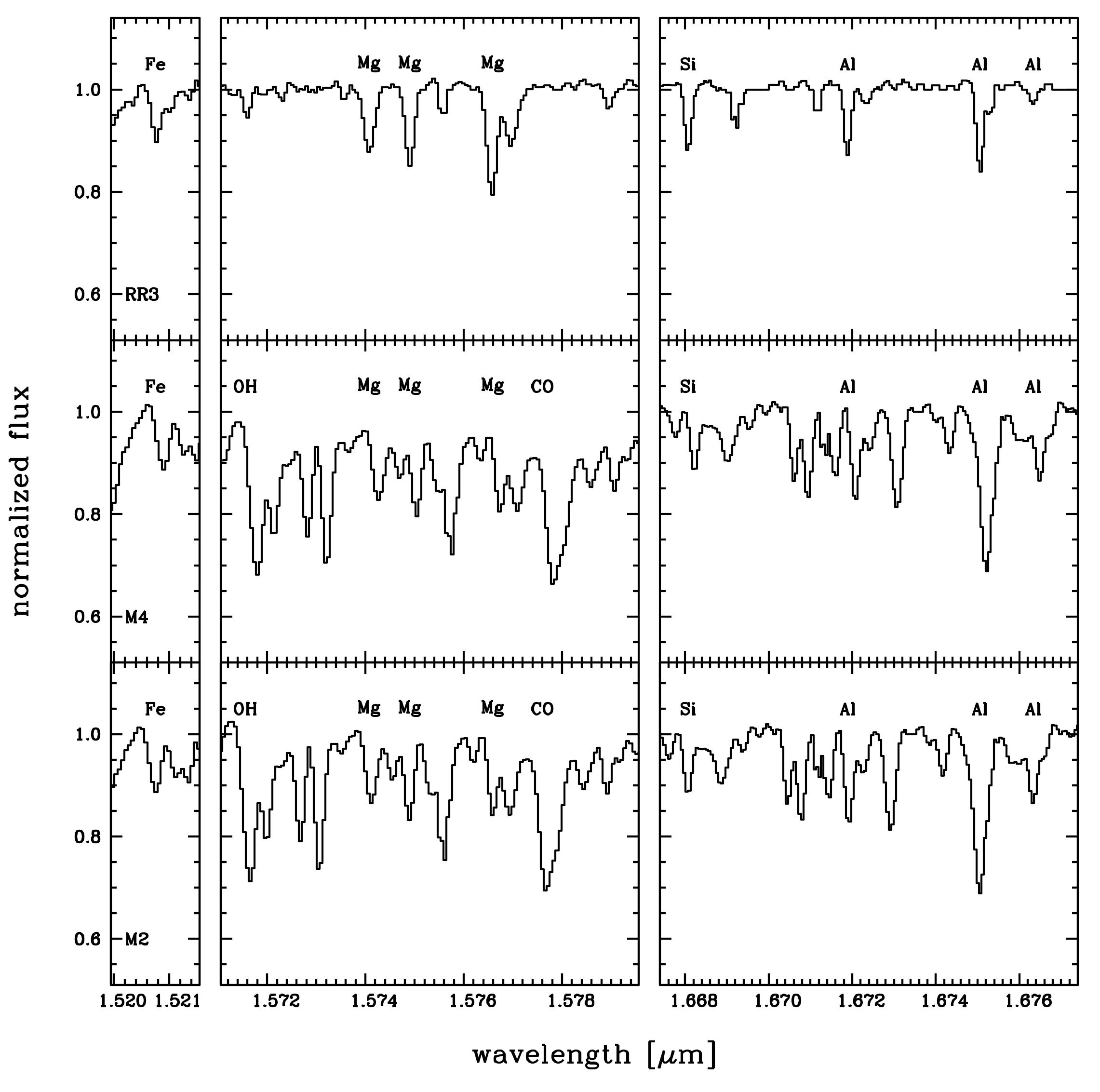}
\caption{Portions of the observed spectra of the M2, M4 and RR3 variable stars of Ter5 around some features of interest.}
\label{spectra}
\end{center}
\end{figure*}  

\section{Spectroscopic observations and data analysis}
\label{synth}

We used X-SHOOTER \citep{ver11} at the VLT under program 097.D-0337
(PI: L. Origlia), to observe the three most isolated RR Lyrae (namely,
RR1, RR2, RR3)\footnote{RR4, instead, has several neighbors that
  complicate its observation with medium-high resolution,
  seeing-limited spectrographs.}  and the six detected Miras (see
Table~\ref{var}).  We selected the VIS and near IR ARMs to
simultaneously acquire spectra in the Calcium triplet region with the
$0.7\arcsec$ slit at R$\simeq$11,000, and in the JHK bands with the
$0.6\arcsec$ slit at R$\simeq$8,000. This allowed us to measure
several atomic lines and, in the cool Miras, also molecular CO and OH
lines, from which deriving radial velocities (RVs) and chemical
abundances.

The acquisition of X-SHOOTER spectra has been performed by nodding on
slit, with a typical throw of a few arcsec, for an optimal subtraction
of the background and the detector noise.  The reduction of the
X-SHOOTER spectra has been performed by using the ESO X-SHOOTER
pipeline version 3.1.0 to obtain 2D rectified and wavelength
calibrated spectra.  Order and 1D spectrum extraction has been
performed manually.  Total on-source exposure times were 11 min on
Mira variables and 40 min on RR Lyrae.  Overall signal-to-noise ratios
of 30-50 per resolution element have been measured on the final
spectra.

For spectral analysis we have used the MARCS model atmospheres
\citep{gus08} and the code described in detail in \citet{ori02} and
\citet{ori04}, already used to compute synthetic spectra for normal
giant stars in Ter5 \citep{ori11,ori13}.  The code uses the LTE
approximation and it includes thousands of near IR atomic transitions
from the Kurucz
database\footnote{http://www.cfa.harvard.edu/amp/ampdata/kurucz23/sekur.html},
\citet{bie73}, and \citet{mel99}, while molecular data are taken from
our \citep[][ and subsequent updates]{ori93, ori97} and B. Plez
(private communications) compilations.  We use the \citet{gv98}
abundances for the Solar reference.  A list of suitable lines for each
measurable chemical element, free from significant blending and/or
contamination by telluric absorption and without strong wings, has
been identified.  Chemical abundances have been derived by minimizing
the scatter between observed and synthetic spectra with suitable
photospheric parameters and also using as a figure of merit equivalent
width measurements of selected lines.  The typical random error of the
measured line equivalent widths is 20-30 m\AA, mostly arising from a
$\pm $1-2\% uncertainty in the placement of the pseudo-continuum, as
estimated by overlapping the synthetic and the observed spectra.  This
error corresponds to abundance variations of about 0.1 dex, comparable
with the typical 1$\sigma $ scatter ($\le$0.15 dex) in the derived
abundances from different lines.  The errors quoted in Table~\ref{var}
for the final abundances were obtained by dividing the 1$\sigma $
scatter by the square root of the number of used lines, typically a
few per species.  When only one line was available, we assumed a
0.15~dex error.

\section{Results}

Neutral atomic lines in the H-band have been used to derive abundances
of Fe, Ca, Si, Mg, Ti and Al.  Additional atomic lines of Mg, Al, Na,
K, and Mn in the J-band and of Ti, Al, and Na in the K-band have been
also used to derive abundances for the corresponding metals.  These
lines have been also used to derive heliocentric RVs.  OH lines in the
spectra of the cool Miras have been used to derive abundances of O in
those variables.  Detailed information on the best-fit estimates of
the stellar parameters, final RVs and chemical abundances for the
observed variables are given in the next two sub-sections.

\subsection{RR Lyrae variables}
\label{RR}

\begin{figure*}
\begin{center}
\includegraphics[scale=0.18]{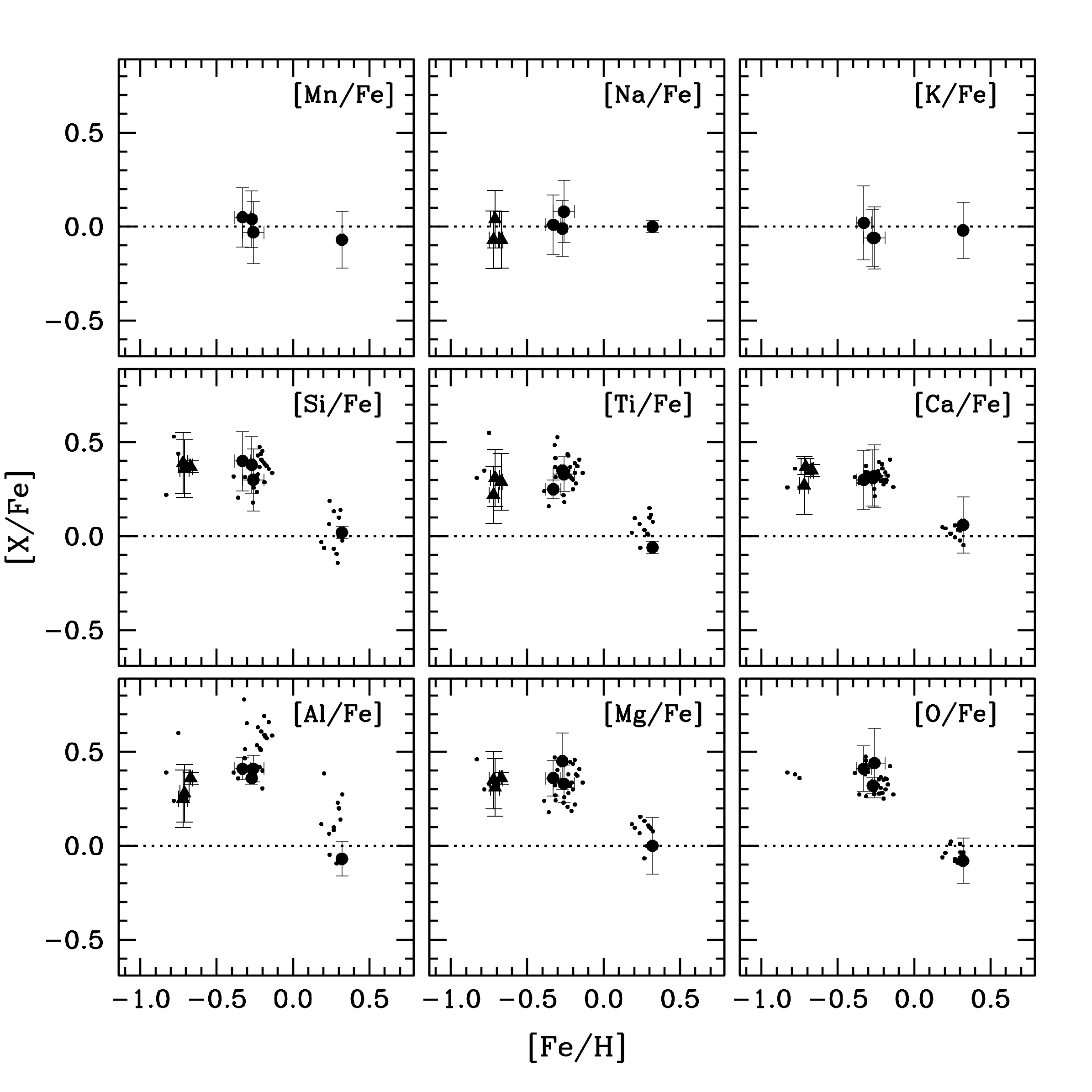}
\caption{Abundance ratios [X/Fe] as a function of [Fe/H] for the
  member RR Lyrae (filled triangles), Mira variables (large circles)
  and normal RGB stars (small dots, from \citealp{ori11,ori13}) of
  Ter5.}
\label{ratios}
\end{center}
\end{figure*}  

The three observed RR Lyrae stars have quite long periods.  Although,
these are typical of metal poor systems \citep{fio15}, they may also
be detected in systems with a red horizontal branch morphology, as it
is observed in the metal rich SPs of the bulge and its globular
clusters \citep[e.g.][and references therein]{kun16,kun18}.  

Given that the stellar parameters of the RR Lyrae stars vary with the pulsation period, 
we have computed theoretical models \citep[see][for details]{mar13} for the inferred long periods and for metallicities 
between solar and one tenth solar, and predict corresponding RV, effective temperature ($\rm T_{eff}$) and gravity (log~g) variability curves. 
We have found maximum amplitudes of about $\pm 20$ km s$^{-1}$ in RV with respect to the average one, 600 K in $\rm T_{eff}$, and 0.5 dex in log~g. 
These predictions, coupled with the spectroscopic observation epoch, allow us to provide first guess RV, $\rm T_{eff}$ and log~g. 
We thus computed a grid of synthethic spectra with varying abundances and stellar parameters  according to the variability curves.
Best-fit estimates from spectral synthesis turned out to be in excellent agreement with those predicted by the variability curves 
at the epoch of the spectroscopic observation and fully consistent with those expected for stars in the instability strip.

An average microturbulence velocity of 2 km s$^{-1}$ and log~g
of 3.0 dex have been assumed for all the stars, while the adopted
temperatures ($\ge$ 6000 K) are reported in Table~\ref{var}.
Systematic uncertainties of $\pm$200 K in $\rm T_{eff}$, $\pm$0.5 in
log~g and $\pm 0.5$ km s$^{-1}$ in microturbulence velocity have been
considered, and their impact on individual abundances turns out to be
$\le$0.1 dex, while the abundance ratios are practically unaffected.

The systemic velocity of Ter5 is $\approx -83$ km s$^{-1}$ and its
velocity dispersion in the central region is $\approx 15$ km
s$^{-1}$ and its \citep{mas14b}.  Hence. according to the measured RVs
(see Table~\ref{var}), all the three observed RR Lyrae are fully
consistent with being Ter5 members.

The derived chemical abundances and abundance ratios of Fe, Ca, Si,
Mg, Ti, Al, and Na are listed in Table~\ref{var} and plotted in
Fig.\ref{ratios}.  An iron abundance [Fe/H]$\sim -0.7$ dex, enhanced
[$\alpha$/Fe] and [Al/Fe], and solar-scaled [Na/Fe] have been
homogeneously inferred for the three RR Lyrae.  

This metallicity, together with the best-fit periods and the average V
magnitudes, can be also used to constrain the distance of Ter5, by
means of the period-luminosity relation by \citet{catelan04}.  V
magnitudes in the Johnson-Cousin system have been obtained from the
mean magnitudes in the HST filters and photometric transformations
computed by using the stars in common between the WFPC2 data-set used
in the present work and the ACS catalog published by \citet{fer09}.
We found V magnitudes of 21.77, 22.59 and 21.83 for RR1, RR2 and RR3,
respectively.  The proper extinction value was then associated to
each candidate RR Lyrae star by using the differential reddening map
derived by \citet{mas12}.  We obtain a distance d=$6.6^{+2.0}_{-1.6}$
kpc, $5.9^{+1.9}_{-1.5}$ kpc and $5.9^{+2.0}_{-1.5}$ kpc for RR1, RR2
and RR3 respectively.  The errors are obtained by assuming a $10\%$
uncertainty on the adopted E(B-V) values.  The derived values well
match the distance of 5.9$\pm$0.5 kpc obtained by \citet{val07} from
IR photometry of the Red Giant Branch (RGB) of Ter5.

Interestingly, the period-metallicity distribution of the Ter5 RR Lyrae stars resembles that of the RR Lyrae 
in the bulge globular clusters NGC 6388 and NGC 6441 \citep[see e.g.][and references therein]{pri00}.
However, the uncertainty in the amplitude estimates from the current light curves prevent us 
to use their period-amplitude distribution as a diagnostic tool for checking 
their possible association with a population of super-luminous, He-rich stars,
as suggested for NGC 6388 and NGC 6441 \citep{rich97,pri02,bus07,bro16,tai17}.

\subsection{Mira variables}
\label{mira}
Very few spectroscopic studies exist on Mira variables in general, and
most of them have been focused on the complex kinematics of their
atmospheres \citep[e.g.][]{hink78,hink82,leb05,wit11} and occasionally
on the determination of C/O abundance ratios
\citep[e.g.][]{leb14,hink16} from molecular CO and OH lines in near IR
spectra.  However, very recently and for the first time, some iron,
$\alpha$-element and sodium abundances of a Mira star in the globular
cluster NGC 5927 have been measured by \citet{dor18}, using J-band
spectroscopy.

Mira variables have complex atmospheres, characterized by
sub-structures with different [low] temperatures, gravities and
velocity fields, that also change with the pulsational phase.  A cool
photosphere with temperatures as low as $\approx$3000 K is believed to
be the source of the atomic lines, as well as of most of the molecular
CO and OH features.  However, especially near the maximum, a
$\approx$1000 K gaseous component likely in the inner portion of a
circumstellar shell, can contribute to low-excitation molecular
absorption lines.  Moreover, velocity gradients, departure from LTE in
the outer layers and/or P Cygni type emission arising in a
circumstellar shell might weaken some absorption lines.  Although
hydrostatic model atmospheres cannot satisfactory reproduce all the
observed features, especially near the maximum, they can still be used
to perform some chemical abundance analysis from high excitation
atomic and molecular lines that originate in the innermost region of
the Mira's photosphere.

The six candidate Miras towards Ter5 have been observed at a random
phase far from the maximum (see Table~1).  Effective temperatures have
been spectroscopically determined from X-SHOOTER spectra, by computing
the CO indices in correspondence of the first-overtone (2-0) and (3-1)
bandheads in the K-band and using the calibrations reported in
\citet{schu16}.  We find values in the 3000-3400 K range (see
Table~\ref{var}), consistent with an observation epoch far from the
light curve maximum.  A microturbulence velocity of 2 km s$^{-1}$ and
a surface gravity log~g of 0.5 dex have been adopted, consistent with
the typical values measured in Ter5 non-variable cool giants near the
RGB tip \citep{ori11,ori13}.  Systematic uncertainties of $\pm$200 K
in T$_{eff}$, $\pm$0.5 in log~g and $\pm 0.5$ km s$^{-1}$ in
microturbulence velocity imply abundance variations between 0.1 and
0.2 dex.

Second overtone molecular band-heads of $^{12}$CO have been routinely
used to derive carbon abundances in normal RGB stars.  However, in
pulsating Miras, these band-heads can be affected by kinematics, thus
making difficult to disentangle abundance from velocity gradient
effects. Hence, we did not attempt to obtain any carbon abundance
estimate from the spectral synthesis.

According to the measured RVs (see Table~\ref{var}), only the
innermost Miras (M1 to M4) are consistent with being Ter5 members,
also in agreement with the proper motions measured by Gaia.  M4 has
been observed after its minimum of luminosity, hence its RV is
expected to be negative with respect to the mean.  The measured value
of $-119$ km s$^{-1}$, although significantly more negative than the
other Miras and the systemic velocity, is still consistent with a Ter5
membership at $\approx 1.5-2.0\sigma$ level.  Interestingly,
\citet{mat05} detected SiO maser emission of M4 at V(LSR)$=-106$ km
s$^{-1}$, which corresponds to $-96$ km s$^{-1}$, giving an
additional, strong support for the membership of this star.

The derived chemical abundances and abundance ratios of Fe, Ca, Si,
Mg, Ti, Al, Na, Mn, K and O are listed in Table~\ref{var} and plotted
in Fig.\ref{ratios}.  An iron abundance [Fe/H]$\sim-0.3$ dex, enhanced
[$\alpha$/Fe] and [Al/Fe], and about solar-scaled [Mn/Fe], [Na/Fe] and
K[Fe] have been inferred for the three Miras M1, M2, and M3 with
P$<$300 d, while M4 and the other two Miras with longer periods
(P$>$300 d) have super-solar iron ([Fe/H]$\sim +0.3$ dex) and about
  solar-scaled [$\alpha$/Fe], [Al/Fe], [Na/Fe], [Mn/Fe], and [K/Fe].

\section{Discussion and Conclusions}
The seven variables (three RR Lyrae and four Miras) that have been
found to have RVs consistent with being members of Ter5 have very
different metallicities and [$\alpha$/Fe] abundance ratios.  The
agreement with the earlier non-variable star studies of M giants in
Ter5 \citep{ori11,ori13} is striking.  The three RR Lyrae show
[Fe/H]$\sim-0.7$ dex and enhanced [$\alpha$/Fe]$\sim$+0.3 dex, nicely
matching the values obtained for the most metal poor population
detected in Ter5 \citep[][see also Fig.~\ref{ratios}]{ori13}.  The
three Miras with P$<$300 d (namely, M1, M2, and M3) appear to be
clumped at [Fe/H]$\sim-0.3$ dex and [$\alpha$/Fe]$\sim$+0.3 dex, again
nicely matching the values measured in the dominant sub-solar
population of Ter5.  The exceptionally long-period Mira M4 at
[Fe/H]=+0.32 and solar-scaled [$\alpha$/Fe] is fully consistent with
the super-solar component that \citet{fer16} found to be 7 Gyrs
younger than the metal-poor SPs of Ter5.  Consistently, the mass of
the M4 Mira is expected to be significantly larger than the mass of
the other Ter5 Miras (M1, M2 and M3) at lower metallicity and with
significantly shorter periods. Indeed, we note that an increase in the
He content and/or a super-solar metallicity alone are not sufficient
to explain the long period of M4.  By using equation (15) in
\citet{fea96} we obtain a current mass of $\sim 0.5 M_{\odot}$ for M1,
M2 and M3 and $\sim 1 M_{\odot}$ for M4, that is a difference in mass
of $\Delta M \sim 0.5 M_{\odot}$. This value is in nice agreement with
the mass difference $\Delta M_{\rm TO} \sim 0.4 M_{\odot}$ that
\citet{fer16} measured at the main sequence Turn Off for the two
sub-populations of Ter5 (i.e., $M_{\rm TO}=0.92$ for the 12 Gyr old
and sub-solar metallicity component, and 1.32 $M_{\odot}$ for the 4.5
Gyr old and super-solar metallicity one).

Interestingly, the other two long period Miras, namely M5 and M6,
which are likely bulge field stars, have abundances and abundance
patterns similar to those of M4, providing an additional evidence of
the connection between Ter5 and the bulge, and of young ages for some
stars in both systems.

Finally, our finding that both the RR Lyrae and Mira variables in Ter5
have metallicities and ages consistent with those of non-variable
stars, is a confirmation of the scenario proposed in \citet[][and
  references therein]{fer16}, where Ter5 experienced a complex
evolutionary history and is currently comprised of sub-populations
with mutiple and discrete ages and metallicities.

\acknowledgements ED aknowledges support from The Leverhulme Trust Visiting Professorship Programme VP2-2017-030. 
RMR acknowledges support from grants AST-1413755, 1518271 from the National Science Foundation.  DM acknowledges
financial support from a Vici grant from NWO.  This research is part
of the Cosmic-Lab Project at the Bologna University.


\begin{thebibliography}{} 
\bibitem[Barbuy et al.(1998)]{bar98} 
Barbuy, B., Bica, E., \& Ortolani, S.\ 1998, \aap, 333, 117
\bibitem[Barning(1963)]{barning63} Barning, F.~J.~M.\ 1963, \bain, 17, 22
\bibitem[Bekki \& Freeman(2003)]{bekki03} 
Bekki, K., \& Freeman, K.~C.\ 2003, \mnras, 346, L11
\bibitem[Bekki \& Norris(2006)]{bekki06} 
Bekki, K.,  \& Norris, J. E., 2006, \apj, 637,109
\bibitem[Bensby et al.(2013)]{ben13}
Bensby, T. et al. 2013, \aap, 549, 147
\bibitem[Behrendi, Burkert \& Schartmann(2016)]{beh16}
Behrendi,M., Burkert, A., \& Schartmann, M., 2016, \apj, 819, 2
\bibitem[Bi\`emont \& Grevesse(1973)]{bie73}
Bi\`emont, E., \& Grevesse, N. 1973, {\it Atomic Data and Nuclear Data
Tables}, 12, 221
\bibitem[Brown et al.(2016)]{bro16}
Brown, T.M., et al., 2016, \apj, 822, 44
\bibitem[Busso et al.(2007)]{bus07}
Busso, G., et al., 2007, \aap, 474, 105
\bibitem[Cadelano et al.(2018)]{cad18}
Cadelano, M., et al., 2018, \apj, 855, 125
\bibitem[Catchpole et al.(2016)]{cat16}
Catchpole, R.M., Whitelock, P.A., Feast, M.W., Hughes, S.M.G., Irwin, M., Alard, C., 2016, MNRAS, 455, 2216
\bibitem[Carollo et al.(2007)]{car07}
Carollo, C. M., Scarlata, C., Stiavelli, M., Wyse, R.F.G., \& Mayer, L., 2007, \apj, 658, 960
\bibitem[Catelan et al.(2004)]{catelan04} Catelan, M., Pritzl, B.~J., \& Smith, H.~A.\ 2004, \apjs, 154, 633
\bibitem[Clement et al.(2001)]{cle01}
Clement, C.M., et al., 2001, AJ, 122, 2587
\bibitem[Dalessandro et al.(2018)]{dal18}
Dalessandro, E., et al., 2018a, \apj, 859, 15
\bibitem[Dolphin(2000)]{dolphin00} Dolphin, A.~E.\ 2000, \pasp, 112, 1397 
\bibitem[D'Orazi et al.(2018)]{dor18}
D'Orazi, V, et al., 2018, \apj, 855, 9
\bibitem[Edmonds et al.(2001)]{edm01}
Edmonds,  P.D., Grindlay, J.E., Cohn, H., \& Lugger, P., 2001, \apj, 547, 829 
\bibitem[Elmegreen, Bournaud \& Elmegreen(2008)]{elm08}
Elmegreen, B.G., Bournaud, F., \& Elmegreen, D.M., 2008, \apj, 688, 77
\bibitem[Feast(1996)]{fea96} 
Feast, M.W., 1996, \mnras, 278,11 
\bibitem[Ferraro et al.(2009)]{fer09} 
Ferraro, F.R., et al., 2009, Nature, 462, 483
\bibitem[Ferraro et al.(2015)]{fer15} 
Ferraro, F.R., Pallanca, C., Lanzoni, B., Cadelano, M., Massari, D., Dalessandro, E, Mucciarelli, A., 2015, \apj, 807, L1
\bibitem[Ferraro et al.(2016)]{fer16} 
Ferraro, F.R., Massari, D., Dalessandro, E., Lanzoni, B., Origlia, L., Rich, R.M., \& Mucciarelli, A., 2016, \apj, 828, 75
\bibitem[Fiorentino et al.(2012)]{fio12}
Fiorentino, G., Contreras Ramos, R., Tolstoy, E., Clementini, G., Saha, A., 2012, \aap, 539, 138
\bibitem[Fiorentino et al.(2015)]{fio15}
Fiorentino, G., et al., 2015, \apj, 798, L12
\bibitem[Gaia Collaboration et al.(2018)]{gaia18} 
Gaia Collaboration, Brown, A.~G.~A., Vallenari, A., et al.\ 2018, \aap, 616, A1 
\bibitem[Genzel et al.(2011)]{gen11}
Genzel, R., et al., 2011, \apj, 733, 101
\bibitem[Gonzalez et al.(2015)]{gon15}
Gonzalez, O., et al., 2015, \aap, 584, 46
\bibitem[Grevesse \& Sauval(1998)]{gv98} 
Grevesse, N., \& Sauval, A. J. 1998, {\em Space Science Reviews}, 85, 161
\bibitem[Gustafsson et al.(2008)]{gus08}
Gustafsson B., Edvardsson B., Eriksson K., Joergensen U.G., Nordlund A., Plez B., 2008, \aap, 486, 951
\bibitem[Hill et al.(2011)]{hil11}
Hill, V., Lecureur, A., Gomez, A., Zoccali, M., Schultheis, M., Babusiaux, C., Royer, F.,
Barbuy, B., Arenou, F., Minniti, D., \& Ortolani, S. 2011, \aap, 535, 80
\bibitem[Hinkle(1978)]{hink78}
Hinkle, K.H., 1978, \apj, 220, 210
\bibitem[Hinkle, Hall, \& Ridgway(1982)]{hink82}
Hinkle, K.H., Hall, D.B., \& Ridgway, S.T., 1982, \apj, 252, 697
\bibitem[Hinkle, Lebzelter, \& Straniero(2016)]{hink16}
Hinkle, K.H., Lebzelter, T., \& Straniero, O., 2016, \apj, 825, 38
\bibitem[Holtzman et al.(1995)]{holtzman95} Holtzman, J.~A., Burrows, C.~J., Casertano, S., et al.\ 1995, \pasp, 107, 1065
\bibitem[Immeli et al.(2004)]{imm04} Immeli, A., Samland, M., Gerhard,
  O., \& Westera, P.\ 2004, \aap, 413, 547
\bibitem[Johnson et al. (2011)]{joh11}
Johnson, C.I., Rich, R.M., Fulbright, J.P, Valenti, E., \& McWilliam, A. 2011, \apj, 732, 108
\bibitem[Johnson et al.(2014)]{joh14}
Johnson, C.I., Rich, R.M., Kobayashi, C., Kunder, A., \& Kock, A., 2014, \aj, 148, 67
\bibitem[J\"onsson et al.(2017)]{jon17}
J\"onsson, H., Ryde, N., Schultheis, M., \& Zoccali, M., 2017, \aap, 598, 101
\bibitem[Kunder et al.(2016)]{kun16}
Kunder, A., et al., 2016, \apj, 821, 25 
\bibitem[Kunder et al.(2018)]{kun18}
Kunder, A., et al., 2018, \aj, 155, 171
\bibitem[Lanzoni et al.(2010)]{lan10} 
Lanzoni, B., et al., 2010, \apj, 717, 653
\bibitem[Lebzelter et al.(2004)]{leb05}
Lebzelter, T., Wood,P.R.,  Hinkle, K.H.,  Joyce, R.R., \& Fekel, F.C., 2005, \aap, 432, 207
\bibitem[Lebzelter et al.(2004)]{leb14}
Lebzelter, T., Wood,P.R.,  Nowotny, W.,  Hinkle, K.H., H\"ofner, S., \& Aringer, B., 2014, \aap, 567, 143
\bibitem[Lomb(1976)]{lomb76} Lomb, N.~R., 1976, \apss, 39, 447 
\bibitem[Marconi et al.(2013)]{mar13}
Marconi, M., Molinaro, R., Ripepi, V., Musella, I., Brocato, E., 2013, \mnras, 428, 2185
\bibitem[Massari et al.(2012)]{mas12}
Massari, D., et al., 2012, \apj, 755, 32
\bibitem[Massari et al.(2014a)]{mas14a}
Massari, D., et al., 2014a, \apj, 791, 101
\bibitem[Massari et al.(2014b)]{mas14b}
Massari, D., et al., 2014b, \apj, 795, 22 
\bibitem[Massari et al.(2015)]{mas15}
Massari, D., et al., 2015, \apj, 810, 69 
\bibitem[Matsunaga et al.(2005)]{mat05}
Matsunaga, N., Deguchi, S., Ita, Y., Tanabe, T., \& Nakada, Y., 2005, PASJ, 57, L1
\bibitem[McKenzie \& Bekki(2018)]{mck18}
McKenzie, M., \& Bekki, K., 2018, \mnras, 479, 3126
\bibitem[Mel\'endez \& Barbuy(1999)]{mel99}
Mel\'endez, J., \& Barbuy, B. 1999, \apjs, 124, 527 
\bibitem[Ness et al.(2013a)]{ness13a}
Ness, M., Freeman, K., Athanassoula, E., Wylie-De-Boer, E., Bland-Hawthorn, J.,
Asplund, M., Lewis, G.F., Yong, D., Lane, R.R., \& Kiss, L.L. 2013a, \mnras, 430, 836
\bibitem[Ness et al.(2013b)]{ness13b}
Ness, M., Freeman, K., Athanassoula, E., Wylie-De-Boer, E., Bland-Hawthorn, J.,
Asplund, M., Lewis, G.F., Yong, D., Lane, R.R., Kiss, L.L., \& Ibata, R. 2013b, \mnras, 432, 2092
\bibitem[Origlia, Moorwood \& Oliva(1993)]{ori93} 
Origlia, L., Moorwood, A. F. M., \& Oliva, E. 1993, \aap, 280, 536
\bibitem[Origlia et al.(1997)]{ori97} 
Origlia, L., Ferraro, F. R., Fusi Pecci, F., \& Oliva, E. 1997, \aap, 321, 859
\bibitem[Origlia, Rich \& Castro(2002)]{ori02} 
Origlia, L., Rich, R. M., \& Castro, S. 2002, \aj, 123, 1559
\bibitem[Origlia \& Rich(2004)]{ori04} 
Origlia, L., \& Rich, R. M. 2004, \aj, 127, 3422
\bibitem[Origlia et al.(2011)]{ori11} 
Origlia, L., et al., 2011, \apj, 726, 20
\bibitem[Origlia et al.(2013)]{ori13} 
Origlia, L., Massari, D., Rich, R.M., Mucciarelli, A., Ferraro, F.R., Dalessandro, E., Lanzoni, B., 2013, \apj, 779, 5
\bibitem[Ortolani, Barbuy, \& Bica(1996)]{ort96}
Ortolani, S., Barbuy, B., \& Bica, E., 1996, \aap, 308, 733
\bibitem[Pritzl et al.(2000)]{pri00}
Pritzl, B., Smith, H. A., Catelan, M., \& Sweigart, A.V., 2000, \apj 530, 41
\bibitem[Pritzl et al.(2002)]{pri02}
Pritzl, B.,J., Smith, H. A., Catelan, M., \& Sweigart, A.V., 2002, \aj, 124, 949
\bibitem[Ransom et al.(2005)]{ransom05} 
Ransom, S.~M., Hessels, J.~W.~T., Stairs, I.~H., Freire, P.~C.~C., Camilo, F., Kaspi, V.~M.,
  \& Kaplan, D.~L.\ 2005, Science, 307, 892
\bibitem[Rich et al.(1997)]{rich97} 
Rich, R.M., et al., 1997, \apj, 484, 25
\bibitem[Rich, Origlia \& Valenti(2012)]{ric12} 
Rich, R.M., Origlia, L.  \& Valenti, E.,  2012, \apj, 746, 59
\bibitem[Rojas-Arriagada et al.(2014)]{roj14}
Rojas-Arriagada, A., et al., 2014, \aap, 569, 103 
\bibitem[Ryde et al.(2016)]{ryd16}
Ryde, N., Schultheis, M., Grieco, V., Matteucci, F., Rich, R. M., \& Uttenthaler, S., 2016, \apj, 831, 40
\bibitem[Schultheis, Ryde \& Nandakumar(2016)]{schu16}
Schultheis, M., Ryde N., \& Nandakumari, G., 2016, \aap, 590, 6
\bibitem[Schultheis et al.(2017)]{sch17}
Schultheis, M. et al., 2017, \aap, 600, 14 
\bibitem[Sloan et al.(2010)]{slo10}
Sloan, G.C., et al., 2010, \apj, 719, 1274
\bibitem[Stetson(1987)]{stetson87} Stetson, P.~B.\ 1987, \pasp, 99, 191
\bibitem[Stetson(1994)]{stetson94} Stetson, P.~B.\ 1994, \pasp, 106, 250 
\bibitem[Tacchella et al.(2015)]{tac15}
Tacchella, S., et al., 2015, \apj, 802, 101
\bibitem[Tailo et al.(2017)]{tai17}
Tailo, M., et al., 2017, \mnras, 465, 1046
\bibitem[Valenti et al.(2007)]{val07} 
Valenti, E., Ferraro, F. R., \& Origlia, L., 2007, \aj, 133, 1287
\bibitem[Vernet et al.(2011)]{ver11}
Vernet, J., et al., 2011, \aap, 536, 105 
\bibitem[Wittkowski et al.(2011)]{wit11}
Wittkowski, M., et al., 2011, \aap, 532, 7
\bibitem[Wood \& Bessell(1983)]{woo83}
Wood, P.R., \& Bessell, M.S., 1983, \apj, 265, 748
\bibitem[Zoccali et al.(2008)]{zoc08} 
Zoccali, M., Hill, V., Lecureur, A., Barbuy, B., Renzini, A., Minniti, D., G{\'o}mez, A., \& Ortolani, S.\ 2008, \aap, 486, 177
\end{thebibliography}
\end{document}